\documentclass[aps,prl,twocolumn,floatfix,superscriptaddress]{revtex4-1}
\usepackage{graphicx}
\usepackage[T1]{fontenc}
\usepackage[utf8]{inputenc}
\usepackage{hyperref}
\usepackage{blindtext}
\usepackage{amsmath,amssymb,braket}
\usepackage{bm}
\usepackage{siunitx}
\usepackage[version=4]{mhchem}

\renewcommand{\vec}[1]{\boldsymbol{#1}}
\renewcommand{\S}{\vec{S}}
\renewcommand{\k}{\vec{k}}

\renewcommand{\r}{\vec{r}}

\newcommand{\zigzagJTwo}{0.8 J_1}
\newcommand{\zigzagJThree}{0.8 J_1}

\newcommand{\stripyJTwo}{0.4 J_1}
\newcommand{\stripyJThree}{-0.2 J_1}

\newcommand{\figref}[1]{Fig.\ref{#1}}

\newcommand{\ADDSNU}{\affiliation{Department of Physics and Astronomy, Seoul National University, Seoul 08826, Korea}}
\newcommand{\ADDIBS}{\affiliation{Center for Correlated Electron Systems, Institute for Basic Science (IBS), Seoul National University, Seoul 08826, Korea}}
\newcommand{\ADDUOS}{\affiliation{Department of Physics, University of Seoul, Seoul 02504, Korea}}

\begin{document}

\title{Topological magnon bands in the zigzag and stripy phases of antiferromagnetic honeycomb lattice}

\author{Ki Hoon Lee}
\email{kihoonlee@snu.ac.kr}
\ADDIBS
\ADDSNU

\author{Suk Bum Chung}
\email{chung.sukbum@gmail.com}
\ADDIBS
\ADDSNU
\ADDUOS

\author{Kisoo Park}
\ADDIBS
\ADDSNU

\author{Je-Geun Park}
\ADDIBS
\ADDSNU

\begin{abstract}
    We investigated the topological property of magnon bands in the collinear magnetic orders of zigzag and stripy phases for the antiferromagnetic honeycomb lattice and identified Berry curvature and symmetry constraints on the magnon band structure. Different symmetries of both zigzag and stripy phases lead to different topological properties, in particular, the magnon bands of the stripy phase being disentangled with a finite Dzyaloshinskii-Moriya (DM) term with non-zero spin Chern number. This is corroborated by calculating the spin Nernst effect. Our study establishes the existence of the non-trivial magnon band topology for all observed collinear antiferromagnetic honeycomb lattice in the presence of the DM term.
\end{abstract}
\maketitle
    
\emph{Introduction}: Topology has emerged over the past few decades as a long-ignored yet most revolutionary concept in condensed matter physics  \cite{RevModPhys.89.040502}. In particular, the last few years have witnessed intensive searches worldwide for topological phases in electronics systems \cite{hasan_colloquium_2010,qi_topological_2011,ando_topological_2013,bernevig_topological_2013,chiu_classification_2016}. And more recently another race began in search for analogous phases in bosonic systems already with a few notable examples: gapped spin liquid, photonic band gap materials, and magnon insulators \cite{kalmeyer_csl_1987,kitaev_csl_2006,banerjee_csl_2016,do_csl_2017,haldane_photonicTI_2008,wang_photonicTI_2008,wang_photonicTI_2009,zhang_magnonTI_2013}.
The realization that the Bloch theorem holds for both bosonic and electronic band structures has led to a couple of new topological magnon bands too: in pyrochlore lattice \cite{mook_tunable_2016,li_weyl_2016,mook_magnon_2017,su_magnonic_2017,jian_weyl_2017} and in 2D Kagome lattice \cite{mook_magnon_2014,chisnell_topological_2015}.
In the light of the recent experimental realization of monolayer magnetic honeycomb lattice of \ce{NiPS3} and \ce{FePS3} \cite{kuo_exfoliation_2016,lee_ising_2016}, \ce{Cr2Ge2Te6} \cite{gong_discovery_2017}, and \ce{CrI3} \cite{huang_layer_2017}, the topological character of magnon band in honeycomb lattice is not only of theoretical interest but also of experimental relevance.

Previous studies of magnon band topology for the honeycomb lattice were restricted to the ferromagnetic and N\'{e}el phases. 
It has been shown that 
the former exhibits the magnon band structure with Dirac points occurring at each valley point \cite{fransson_magnon_2016,pershoguba_dirac_2017}, which 
is gapped out by the next-nearest-neighbor DM interaction producing the magnon analogue of the quantum anomalous Hall (QAH) phase \cite{kim_realization_2016,owerre_first_2016}.
However, the finite DM interaction added to the latter phase, while giving rise to the spin Nernst effect, does not yield a topological magnon band structure  \cite{zyuzin_magnon_2016,cheng_spin_2016}. In addition, 
for the $S_z$ conserving phases, another possible topological phase is the magnon analogue of the quantum spin Hall (QSH) phase arising from the well-defined spin Chern number \cite{yang_qsh_2011}.  
To the best of our knowledge, no simple model is currently available for this topological phase; models proposed so far for the magnon analogue of the QSH include either the bilayer honeycomb with antiferromagnetic interlayer coupling \cite{zyuzin_magnon_2016,owerre_magnon_2016} or 
dipolar interaction \cite{wang_anomalous_2017} with the Aharonov-Casher effect on magnon bands under external electric field for square lattice \cite{nakata_magnonic_2017-1}.
This motivates us to study the magnon topology of other physically feasible collinear spin ordered phases in the honeycomb spin Hamiltonian: the zigzag and stripy phases.

In this Letter, we examined the topological properties of these two phases on the monolayer honeycomb lattice. Our study finds that for both phases the non-trivial Berry phase and Dirac magnon point are protected by spatial (glide-)mirror symmetry. 
We showed that with the DM interaction the stripy phase hosts the magnon analogue of the QSH phase while the zigzag phase features a line-nodal magnon band degeneracy protected by the combination of the non-symmorphic symmetry and the time reversal symmetry.
We demonstrated that the resulting band topology is that of the $\mathcal{C}_S = 1$ QSH, not the $\mathcal{C} = 1$ QAH,
by computing its spin Berry curvature and the edge states for a finite width lattice. 
By contrast, we found that the zigzag phase has a nodal line protected by the non-symmorphic symmetry combined with the time reversal symmetry. 
For both phases, we also calculated the spin Nernst effect, which is the manifestation of the non-trivial topology and the direct consequence of the non-trivial magnon Berry curvature. 
In the remainder of the Letter, we will first explain our spin model for the monolayer honeycomb lattice and the method we used to calculate the magnon band, before presenting our results for the zigzag and stripy phases. 

\emph{Model}: We consider a $J_1$-$J_2$-$J_3$ model for the honeycomb lattice with the next-nearest-neighbor DMI,
\begin{equation}
H_0 = \sum_{n=1}^3 J_n \sum_{\langle i,j \rangle_n } \S_i \cdot \S_j + J_{\textrm{DM}} \sum_{\langle i,j \rangle_2} \nu_{ij} \hat{\vec{z}} \cdot \left(\S_i \times \S_j \right)
\label{eq:Hamiltonian}
\end{equation}
, where $\S_i$ is spin at site $i$ with size $S$ and $\langle i,j \rangle_n$ is the set of pair of $n$th nearest neighbors: 
$\nu_{ij} =\textrm{sign}\sum_{\langle i,k \rangle_1,\langle k,j \rangle_1 }\hat{\vec{z}}\cdot \r_{ik}\times \r_{kj}$, where $\r_{ij} = \r_i - \r_j$ and $\r_i$ is the coordinate of $i$th site.
Note that there is only one $k$ simultaneously satisfying both conditions $\langle i,k\rangle_1 $ and $\langle k,j\rangle_1 $ for given second nearest neighbor pair $\langle i,j\rangle$.
While our analysis is intended to be of general applicability, the Hamiltonian is mostly relevant to 
the single-layer Van der Waals magnetic material family $\textrm{MAX}_3$, such as \ce{NiPS3} and \ce{FePS3} \cite{0953-8984-28-30-301001, kuo_exfoliation_2016,lee_ising_2016}. 
In the phase diagram of $J_1,J_2$ and $J_3$, several ground states were identified such as FM, N\'eel, zigzag and stripy phases while other non-collinear phases can also appear for both cases of $J_1>0$ and $J_1<0$ \cite{fouet_investigation_2001}; for instance, the zigzag phase for $J_1>0$ appears within $J_2/J_1, J_3/J_1 > 0.5$ and we will examine the stripy phase later in this Letter.

It is important to note that spin-orbit coupling, if not zero, has a crucial consequence on the symmetry of the magnetic Hamiltonian and its ground state.
For a non-zero spin-orbit coupling, the symmetric point group of the magnon Hamiltonian for the magnetic ground state contains operations acting both on spin and lattice.
But if the spin-orbit coupling is small enough, we can ignore some of the spin anisotropic exchange or higher order single ion anisotropy.
Such assumption allows us to reduce Hamiltonian to have symmetry higher than the (magnetic) space group and such a symmetry group is called a spin-space group \cite{brinkman_theory_1966}.
For example, the symmetry of Hamiltonian with Heisenberg interaction alone is described by the product of space group of lattice and spin rotation group, which is of higher symmetry than the case with finite spin-orbit coupling.

\emph{Method}: We study the magnon bands of zigzag and stripy phases using the linear spin wave theory (LSWT).
We take the $z$ direction to be the easy axis, which is relevant to the anisotropic DM interaction.
Applying Holstein-Primarkoff (HP) transformation to spin $\vec{\tilde{S}}_i$ in the local spin coordinates taking local magnetization direction as the $z$ direction and $\tilde{S}^{+} \simeq \sqrt{2 S} a, \tilde{S}^{-} \simeq \sqrt{2 S} a^{\dagger}, \tilde{S}_z = S - a^{\dagger}a$, we obtain a quadratic HP boson Hamiltonian in the following form
\begin{align}
H &= \frac{1}{2}\sum_{\alpha,\beta,\k} \psi_{\alpha\k}^{\dagger} H _{\alpha\beta}(\k)
\psi_{\beta\k}\nonumber\\
  &= \frac{1}{2}\sum_{\k,\eta} \left[ E_{\eta}(\k) \gamma_{\eta\k}^{\dagger}
\gamma_{\eta \k} +  E_{\eta}(-\k) \gamma_{\eta, -\k} \gamma_{\eta, -\k}^{\dagger}\right]
\label{eq:HP_Hamiltonian}
\end{align}
, where $\psi_{\alpha\k} = (a_{\alpha,\k}, a_{\alpha, -\k}^{\dagger})^T$ is Nambu spinor of HP boson and $\alpha$, $\beta$ are sublattice index and $\eta$ is band index.
The Hamiltonian can be diagonalized by a para-unitary matrix $T(\k)$ satisfying  
$\sum_{\eta} (\gamma^{\dagger}_{\eta\k},\gamma_{\eta,-\k}) T_{\eta\alpha}^{\dagger}(\k) = \psi_{\alpha\k}^{\dagger}$,
$\sigma_3 = T^{\dagger}(\k) \sigma_3 T(\k) = T(\k) \sigma_3 T^{\dagger}(\k)$,
and
$\sum_{\alpha,\beta} T_{\eta\alpha}^{\dagger}(\k) H_{\alpha\beta}(\k) T_{\beta \eta'}(\k) = \delta_{\eta\eta'}\textrm{Diag}\left( E_{\eta}(\k),E_{\eta}(-\k)\right\}$
where $\sigma_3$ is the Pauli matrix operator acting on particle-hole space \cite{colpa_diagonalization_1978,shindou_topological_2013}; for the explicit forms of Hamiltonians, see our Supplemental Material 
\cite{supplementary}.

\begin{figure}
\includegraphics[width=\columnwidth]{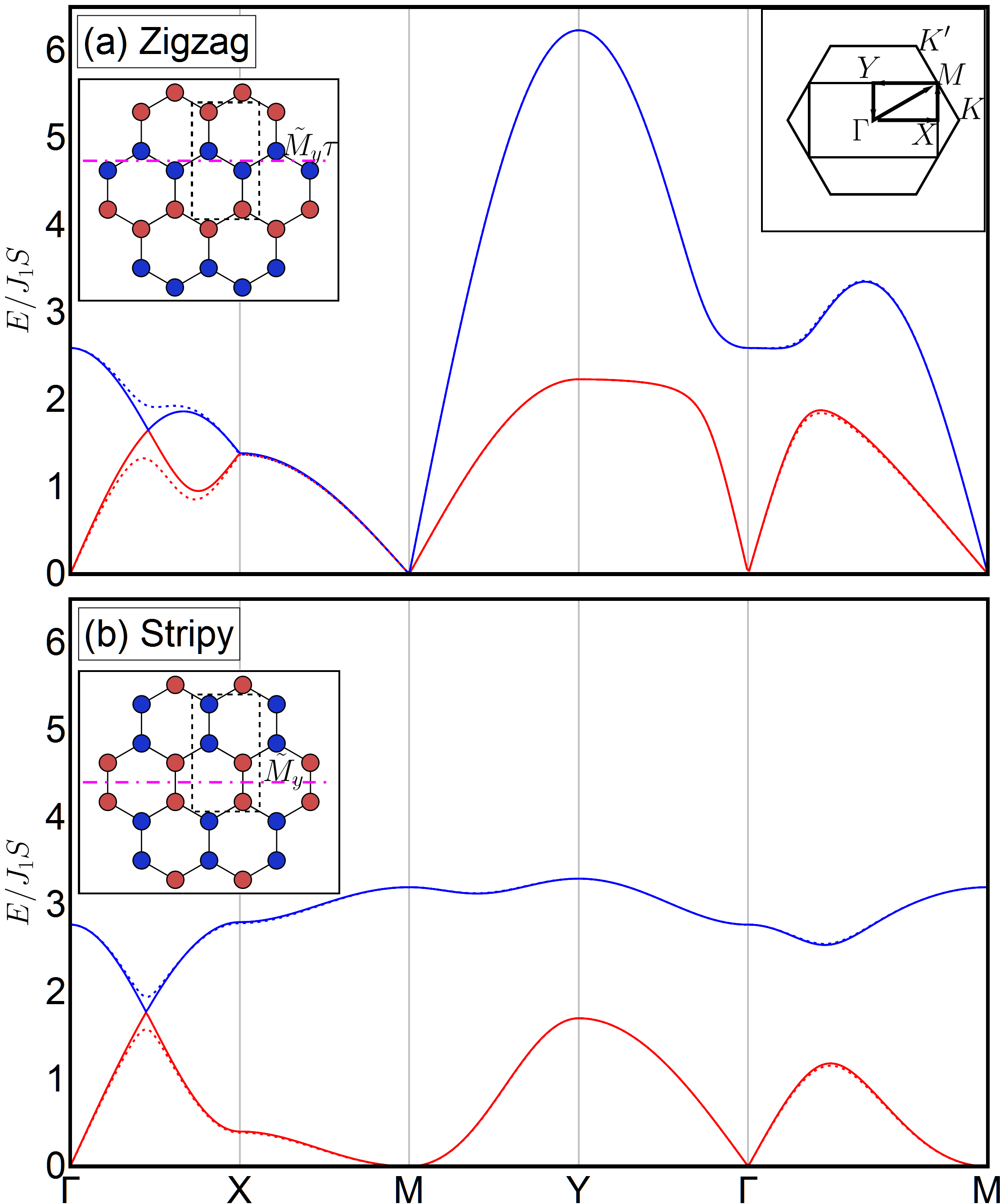}
\caption{Magnon band structure for each phase following paths in the inset of (a). Solid(dashed) line is without(with) DMI.
The figures in insets represent the ground state configurations, and each color represents opposite $s_z$ components.
We can find Dirac magnon on the $\Gamma$-$X$ line, which can be gapped by DMI.
Zigzag phase has four-fold degeneracy on the $X$-$M$ line, which is protected by non-symmorphic symmetry combined with time reversal symmetry.
We used the following parameters for the zigzag phase ($J_2 = \zigzagJTwo, J_3 = \zigzagJThree $) and for the stripy phase  ($J_2 = \stripyJTwo, J_3 = \stripyJThree $).}
\label{fig:band_and_bz}
\end{figure}

\emph{Zigzag phase}: In analyzing the magnon band structure for the zigzag phase, its symmetry properties need to be considered. First, we note that the zigzag phase has the doubled unit cell consisting of four lattice sites and spin configuration as shown in the \figref{fig:band_and_bz} (a) inset.
Second, the symmetry group of Eq.~\eqref{eq:HP_Hamiltonian} composes of elements of the space group of Eq.~\eqref{eq:Hamiltonian} that have the magnetic ground state as its an eigenstate.
The universal symmetry of a collinear phase is $C_{\infty}$ of spin, which is a subgroup of the SO(3) 
of Eq.~\eqref{eq:Hamiltonian} and generated by 
$S_z$.
Therefore, combining with translation symmetry we can assign two eigenvalues to an energy eigenstate, magnetic moment along the magnetization axis $s_z$ and crystal momentum $\k$ in MBZ.
In the presence of an additional symmetry operator $A$ that commute with translation, we will consider two specific cases.
The first case is where the two operators commute, for which we can simply add another eigenvalue to label an energy eigenstate.
The second case is where two operators anti-commute and it guarantees two-fold degeneracy.
To show this, let us assume an energy eigenstate $\ket{s_z,\k}$ labeled by $s_z$ and $\k$. 
The anti-commutation $\{S_z,A\}=0$ guarantees that $A$ should flip the sign of $s_z$, {\it i.e.} $S_z \left(A \ket{s_z,\k}\right) = - A S_z \ket{s_z,\k} = - s_z \left(A \ket{s_z,\k}\right)$, from which we obtain the two-fold degeneracy through $H \ket{-s_z,\k} = H A \ket{s_z,\k} = A H \ket{s_z,\k} = E A \ket{s_z,\k} = E \ket{-s_z,\k}$
(note that if $s_z \neq 0$, $A\ket{s_z,\k} \neq \ket{s_z,\k}$).

In a collinear AFM phase, there are various symmetry operators that combine exchanging sublattices with flipping spins either through two-fold spin rotation or time reversal.
One of such symmetries that do not affect crystal momentum is $C_{2z} \Theta$, where $\Theta$ is the time reversal operation. 
We have $C_{2z}\Theta$ symmetry all in the zigzag, stripy and N\'{e}el phases of the honeycomb lattice.
The two-fold rotation is located at the center of honeycomb for the zigzag and N\'{e}el phases whereas it is at the center of a bonding connecting sites with opposite spin configuration for the stripy phase.
We show below that two relevant symmetries of the zigzag phase that commute with $S_z$ are the glide mirror $\tilde{M}_y \tau_x$, where $\tau_x$ is the half unit cell lattice translation in the $x$-direction (defined in \figref{fig:band_and_bz}) and $\tilde{M}_y e^{i \pi S_x} \tau_x \Theta$; the symmetry operators with the tilde sign here acts only on lattice.

$\tilde{M}_y \tau_x$ protects the two accidental band crossings on the mirror symmetric line $k_y=0$ for $J_{\textrm{DM}}=0$, at which it commutes with the Hamiltonian of Eq.\eqref{eq:Hamiltonian}  \cite{young_dirac_2015}. 
However, for $J_{\textrm{DM}} \neq 0$ as in the N\'{e}el phase, it removes this accidental degeneracy as it breaks $\tilde{M}_y \tau_x$ symmetry due to the pseudo-scalar $\nu_{ij} \equiv \textrm{sign}\sum_{\left<i,k\right>_1,\left<k,j\right>_1}\hat{\vec{z}} \cdot (\vec{r}_{ik} \times \vec{r}_{kj})$ reversing the sign under the mirror operation $\tilde{M}_y$. 
The eigenvalues of the glide mirror $\tilde{M}_y \tau_x$ is $\pm e^{i k_x/2}$ due to  $(\tilde{M}_y \tau_x)^2 = \tau_x^2$, and for $J_{\textrm{DM}}=0$ we can simply assign the same glide mirror eigenvalue to a pair of degenerate bands with opposite $S_z$ eigenvalue as $\tilde{M}_y \tau_x$ commutes with $S_z$.  
The protection for the accidental crossing between two doubly degenerate bands at $J_{\textrm{DM}}=0$ requires their glide mirror eigenvalues being opposite. 
That is exactly the case we found in our work for the zigzag phase, for there is no diagonal term in the representation of $\tilde{M}_y \tau_x$. The sum of mirror eigenvalues of all bands vanishes as $\tilde{M}_y \tau_x$ changes the position of every sublattice. 
While the glide mirror symmetry protects the existing accidental crossings, it does not necessarily guarantee their existence. 
For an accidental crossing to exist on the $\Gamma$-$X$ symmetric line, we need to have $(E_+(\Gamma) - E_-(\Gamma))(\partial_{k_x} E_+(X) - \partial_{k_x} E_-(X))< 0$, where $E_{\pm}$ is the dispersion of bands with $\pm e^{ik_x/2}$ glide mirror eigenvalues. 
This condition arises out of the constraint $E_+(X) = E_-(X)$, which is required by the symmetry that we will now discuss.

At the zone boundary, the glide mirror combined with the time reversal symmetry $g = M_y \tau_x \Theta= \tilde{M}_y e^{i \pi S_x} \tau_x \Theta$ produces the constraint that two bands with opposite glide mirror eigenvalues should be degenerate 
\cite{young_dirac_2015} as long as the zigzag phase is stable.
Not only is $g$ a symmetry, but we have $g^2=-1$ at the MBZ boundary $k_x=\pi$ as $\tau_x^2=e^{i k_x}=-1$ holds there.
An anti-unitary symmetry whose square is $-1$ gives a Kramer-type degeneracy, and - as the operator commutes with $S_z$ - altogether four-fold degeneracy arises, it leads to the sticking of all bands at the zone boundary.
Therefore, at the zone boundary $k_x=\pi$, all four bands are degenerate and so cannot be split \cite{parameswaran_topological_2013}.
It holds even in the presence of the DM interaction as it does not break the symmetry $g$.

\emph{Stripy phase}: The stripy phase has the unit cell identical to that of the zigzag phase with the spin configurations shown in the inset of \figref{fig:band_and_bz} (b). 
Stripy phase has two-fold degeneracy because of degeneracy between states with the opposite 
$s_z$ as in the zigzag phase.
If $J_{\textrm{DM}}=0$, the stripy phase has the spatial mirror symmetry $\tilde{M}_y$, which protects the accidental crossing on mirror invariant line $k_y=0$ for the same reason that the glide mirror symmetry protects that of the zigzag phase. 
The condition to have the crossing is $(E_+(\Gamma) - E_-(\Gamma))(E_+(X) - E_-(X))< 0$, where $E_{\pm}(\k)$ is the dispersion of band with $\pm$ mirror eigenvalue at given momentum $\vec{k}$.

\begin{figure}
\includegraphics[width=\columnwidth]{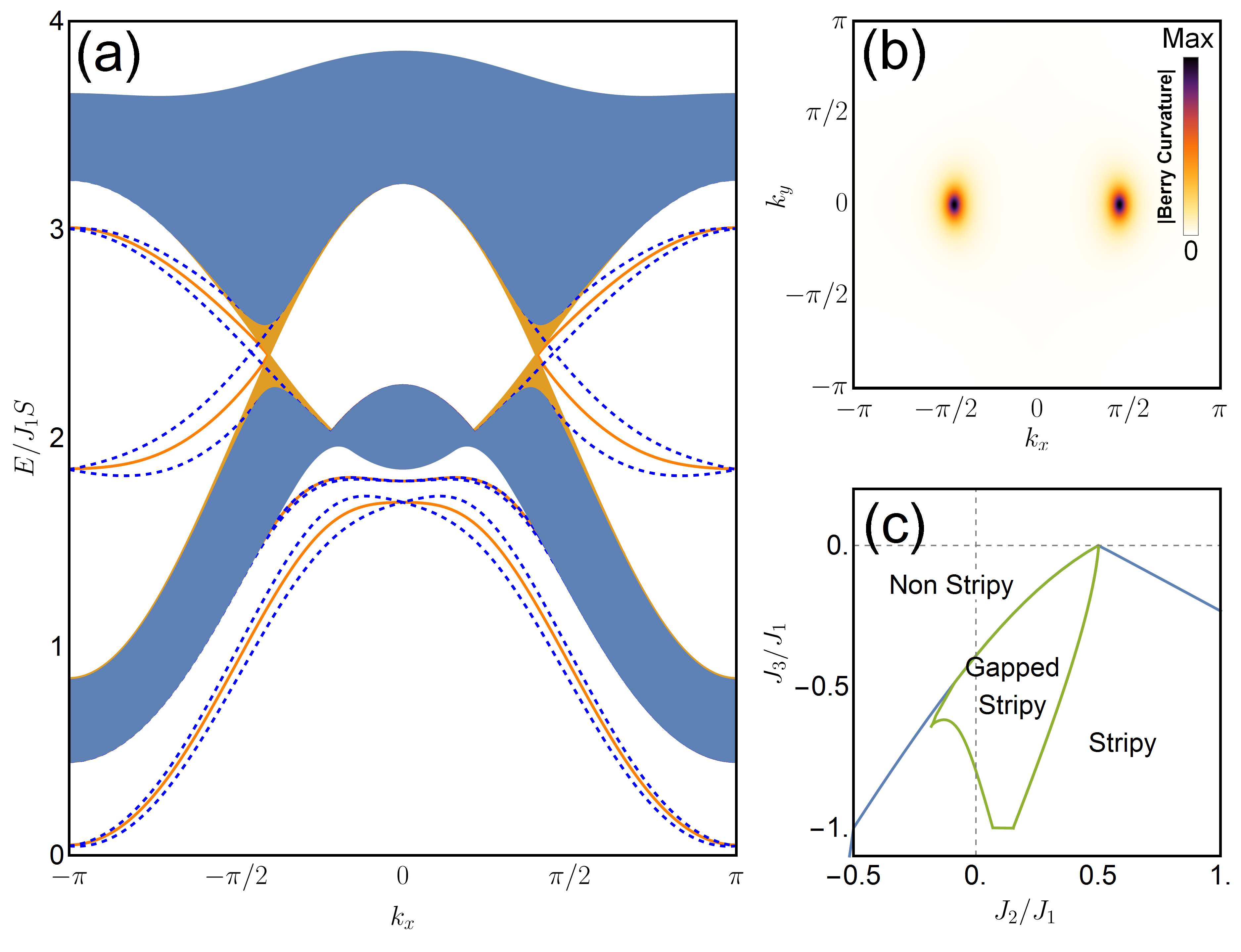}
\caption{(a) Band structure for the stripy phase with finite width along the zigzag direction of honeycomb lattice with DMI (blue) or without DMI (yellow). Lines are the edge states calculated from finite width of 40 magnetic unit cells and colored region represents the bulk bands. To remove the instability from near the edge, we added an easy axis anisotropy term $ 0.3 J_1(1 - S_z^2)$. Other parameters are kept the same as in \figref{fig:band_and_bz}. (b) Berry curvature in Brillouin zone. (c) Gapped stripy phase on top of classical phase diagram of $J_1$-$J_2$-$J_3$ 
for $J_1>0$ (the zigzag phase would require larger positive $J_3/J_1$).
}
\label{fig:edge_mode_berry_curvature_and_phase_diagram}
\end{figure}

With the DM interaction, the stripy phase can exhibit the magnon band structure that is both insulator-like and topologically non-trivial. This is because the DM interaction breaks the spatial mirror symmetry $\tilde{M}_y$ and opens the band crossing as in the zigzag phase.
As this occurs without any nonsymmorphic symmetry constraining bands to be degenerate, two bands can be split to make the magnon band structure insulator-like as depicted in \figref{fig:band_and_bz} (b) with a proper choice of parameters.
In such a case, the topology of the lower band is well-defined,  with the topological invariant under the $S_z$ conservation being the spin Chern number.
The Berry curvature of $n$th band is defined as 
$\Omega_n(\k) = i \epsilon_{\mu\nu z} \left[\sigma_3 \partial_{k_\mu} T^{\dagger}_{\k} \sigma_3 \partial_{k_\nu} T_{\k} \right]_{nn}$ \cite{matsumoto_thermal_2014,murakami_thermal_2016}.
In general, a degenerate band will not have well-defined Berry curvature on its own, but we can separate a pair of degenerate bands using $S_z$ conservation.
The Berry curvature of a band with well-defined $S_z$ eigenvalue is shown in \figref{fig:edge_mode_berry_curvature_and_phase_diagram} (b) and we note that Berry curvature is concentrated near the band edge.
Chern number can be defined for each band as $\mathcal{C}_n = \frac{1}{2\pi }\int dk_x dk_y \Omega_n (\k)$.
Spin Chern number of doubly degenerated lower band is $\mathcal{C}_S = (\mathcal{C}_{\uparrow} - \mathcal{C}_{\downarrow})/2$, where $\mathcal{C}_{\uparrow }$($\mathcal{C}_{\downarrow}$) is Chern number of $s_z=1$($s_z=-1$) \cite{zyuzin_magnon_2016,nakata_magnonic_2017}.
Chern number for each is integer and opposite in its sign (i.e., $\mathcal{C}_{\uparrow} = - \mathcal{C}_{\downarrow}$) because two bands are related by an anti-unitary operator $C_{2z} \Theta$. Therefore, $\mathcal{C}_S$ is also quantized to be integer and for stripy phase $\mathcal{C}_S = \textrm{sign}\left(\textrm{J}_{\textrm{DM}}\right)$.
In this sense, the band topology is analogous to that of the QSH insulator discussed in Ref. \cite{yang_qsh_2011}.

It is important to comment that non-trivial topological number of bosonic bands cannot give rise to quantized transverse response as in fermionic systems.
It is because a bosonic band, lacking the Pauli exclusion principle, cannot be filled uniformly, the only possible exception being the low-temperature transverse conductivity divided by temperature dependent bosonic occupation number where the lowest band is flat and well separated from other bands \cite{nakata_magnonic_2017-1}.
While a non-trivial bosonic band topology can give rise to edge states, their spectra need to lie within the bulk energy gap in order for them to make a greater contribution to transport properties. 
The stripy phase can stabilize such a gapped magnon band structure over a finite parameter space at $J_{\textrm{DM}}/J_1=0.05$ as shown in \figref{fig:edge_mode_berry_curvature_and_phase_diagram} (c). 
In this case, we can have an effective edge magnon transport, as the decay of edge modes to the bulk states requires inelastic scatterings due to the edge modes 
being inside the bulk band gap; it is recently reported that such edge transport is more robust against disorder than the bulk transport \cite{ruckriegel_bulk_2017}.

The dispersion of edge states between the lower and upper band of the stripy phase is shown in \figref{fig:edge_mode_berry_curvature_and_phase_diagram} (a).
The edge modes carry the $S_z$ spin as the HP boson Hamiltonian still commutes with $S_z$ in the nano-ribbon geometry.
There are two types of termination possible, and each type determines the center position of edge modes (whether it to appear in $-\pi>k_x>0$ or $0<k_x<\pi$).
We also had to include a term for easy-axis anisotropy for the nano-ribbon geometry calculations to prevent exponentially decaying deviation from appearing near the edges of the stripy configuration.
This instability is naturally expected as the coordination number is reduced near the edges.
We comment that small non-collinearity will introduce a small gap on the edge modes.

\begin{figure}
    \includegraphics[width=\columnwidth]{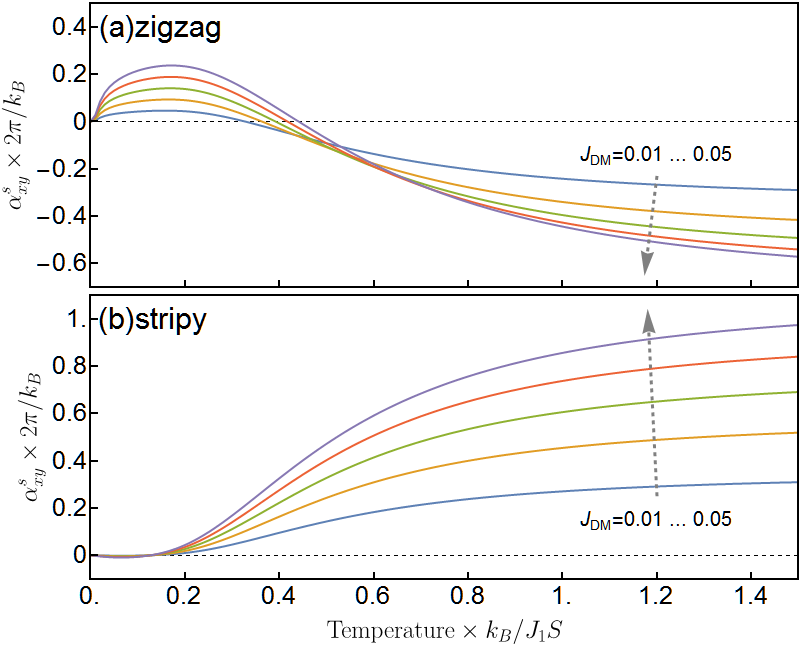}
    \caption{(Color online) Spin Nernst effect in (a) zigzag phase and (b) stripy phase with varying $J_{\textrm{DM}}$}
    \label{fig:SNE}
\end{figure}

Although the spin Berry curvature does not give quantized responses, we still can obtain a finite transverse response.
For example, the FM phase shows a finite thermal Hall effect due to magnons.
Similarly, for the N\'{e}el phase the spin Nernst effect (SNE) was obtained for a finite DM interaction \cite{cheng_spin_2016,zyuzin_magnon_2016}.
The SNE is given by the thermal spin Hall conductivity 
$\alpha_{xy}^s = - \frac{1}{T} \sum_{kn} (S_z \sigma_3)_{nn} \Omega_n(\k) \int_{0}^{E_n ( \k)} d\eta \eta \frac{dg(\eta)}{d\eta}$,
where $g(\eta)$ is the Bose-Einstein distribution function and $n$ is the band index and the band index summation is limited to particle band\cite{zyuzin_magnon_2016}.
The thermal spin Hall conductivity calculated for both zigzag and stripy phases is shown in \figref{fig:SNE}.
We find a sign changing of SNE in the zigzag phase, which can also be found in the N\'{e}el phase \cite{cheng_spin_2016}.
It is due to Berry curvature sign is not constant over the entire bands in the zigzag phase.

A passing comment, both zigzag and stripy phases also appear in extended Kitaev Hamiltonian \cite{chaloupka_kitaev-heisenberg_2010,chaloupka_zigzag_2013,rau_generic_2014,singh_relevance_2012,liu_long_2011}.
But the staggered moment is not out-of-plane, and Kitaev and $\Gamma$ terms break the SO(2) symmetry. 
As a result, neither the spin carried by magnons nor the spin Chern number relying on $S_z$ conservation is well-defined.

\emph{Conclusions}: We have studied the magnon band structure of zigzag and stripy phases for the $J_1$-$J_2$-$J_3$ Heisenberg model with and without DMI for the honeycomb lattice.
In both phases, we found the Dirac magnon, where the accidental crossing is protected by spatial glide mirror or mirror symmetry and the gap can be opened by the DM interaction, but with different topological phases for different symmetries. 
For the zigzag phase, which is commonly found in magnetic systems with honeycomb lattice, 
we found that a nonsymmorphic symmetry combined with time reversal gives rise to the topologically protected line node at zone boundary. 
On the other hand, for the stripy phase, we found that it is possible to realize the magnon analogue of QSH.

\emph{Note added} – While we are preparing our manuscript, a related paper appeared \cite{boyko_evolution_2017}. They have obtained the similar result for the zigzag and stripy phases for the Hamiltonian without DMI.

\begin{acknowledgments}
{\it Acknowledgement}:
We thank SungBin Lee, Cheol-Hwan Park and Bohm Jung Yang for their useful comments. The work at the IBS CCES was supported by the Institute for Basic Science in Korea (IBS-R009-G1, IBS-R009-Y1, and IBS-R009-D1).
\end{acknowledgments}

\bibliography{refs}

\clearpage

\onecolumngrid

	\appendix
	\begin{center}
		{{\bf Supplemental Material for ”Topological magnon bands in the zigzag and str	ipy phases of antiferromagnetic honeycomb lattice”}}
	\end{center}

We present the representations of Holstein-Primarkoff (HP) Hamiltonians for each phase.
To write the representation of HP Hamiltonian, we fix the basis by requiring $S_z$, paramagnetic sublattice index $\tau_3$ in the paramagnetic phase and particle-hole index $\sigma_3$ for a given unit cell to have the following representations:
\begin{align*}
S_z \sigma_3 &: \textrm{Diag}(1,1,1,1,-1,-1,-1,-1)\\
\tau_3 &: \textrm{Diag}(1,1,-1,-1,1,1,-1,-1)\\
\sigma_3 &: \textrm{Diag}(1,-1,1,-1,-1,1,-1,1)
\end{align*}

\begin{figure}[h!]
	\centering{
		\includegraphics[width=70pt]{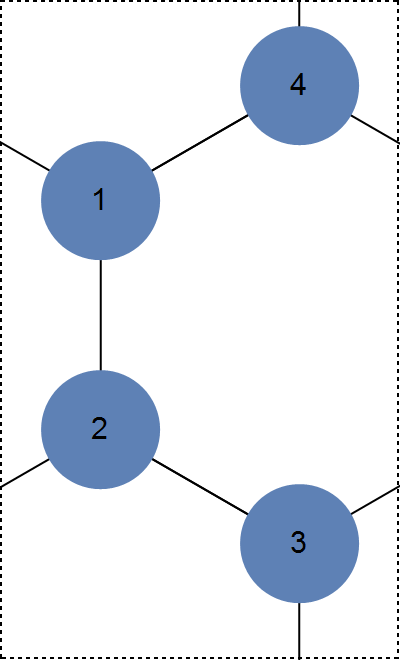}}
	\caption{Convention of the magnetic sublattice index used for this work}
	\label{fig:magnetic_sublattice_index_convention}
\end{figure}
Sign convention for $\tau_3$ is to take $1$ and $3$ to be $+1$ and $2$ and $4$ to be $-1$ in Fig. \ref{fig:magnetic_sublattice_index_convention}.
Note that $S_z$ is defined in the global spin coordinate.
Then, the basis for each phase is
\begin{align}
\textrm{zigzag}:\psi_{\k} &= (b_{1,\k},b^{\dagger}_{3,-\k},b_{4,\k},b^{\dagger}_{2,-\k},b^{\dagger}_{1,-\k},b_{3,-\k},b^{\dagger}_{4,-\k},b_{2,\k})\\
\textrm{stripy}:\psi_{\k} &= (b_{1,\k},b^{\dagger}_{3,-\k},b_{2,\k},b^{\dagger}_{4,-\k},b^{\dagger}_{1,-\k},b_{3,-\k},b^{\dagger}_{2,-\k},b_{4,\k})
\end{align}

The the Hamiltonians are
\begin{align*}
H     &= \frac{1}{2}\sum_{\k} \psi_{\k}^{\dagger} H(\k) \psi_{\k}\\
H(\k) &= \left(
\begin{matrix}
H_{\textrm{I}}(\vec{k}) & 0 \\
0 & H_{\textrm{II}}(\vec{k})
\end{matrix}\right)\\
H_{\textrm{II}}(\vec{k}) &= H_{\textrm{I}}^T(-\vec{k})\\
H_{\textrm{I}}(\vec{k}) &= \left(
\begin{matrix}
A(\vec{k}) & \gamma(\vec{k}) \\
\gamma(\vec{k})^{\dagger} & A(-\vec{k})
\end{matrix}
\right)
\end{align*}

For the zigzag phase,
\begin{align*}
A(\vec{k}) &= 4 \sigma _1 \cos \left(k_y/2\right)
\left(J_{\text{DM}} \sin
\left(k_x/2\right)+J_2 \cos
\left(k_x/2\right)\right)\\
&-\sigma _0 \left(2
J_{\text{DM}} \sin \left(k_x\right)-2 J_2 \cos
\left(k_x\right)+J_1-2 J_2-3 J_3\right)\\
\gamma(\vec{k}) &=e^{-i k_y/3} \left(2 J_1 \sigma _0 e^{i
	k_y/2} \cos \left(k_x/2\right)+\sigma _1
\left(2 J_3 \cos \left(k_x\right)+J_3 e^{i
	k_y}+J_1\right)\right)
\end{align*}
, where $\sigma_i$ is $i$th Pauli matrix.

For the stripy phase,
\begin{align*}
A(\k) &= \sigma _0 \left(2 J_2 \cos \left(k_x\right)+J_1+2 J_2-3 J_3\right) +4 J_2 \sigma _1 \cos \left(k_x/2\right)
\cos \left(k_y/2\right)\\
& +  J_{\text{DM}} \left(4 \sigma _1 \sin \left(k_x/2\right) \cos \left(k_y\right)-2 \sigma _0 \sin
\left(k_x\right)\right) \\
\gamma(\k) &= e^{-i k_y/3} \left(2 J_3 \sigma _0 \cos \left(k_x\right)+2 J_1 \sigma _1 e^{i k_y/2} \cos
\left(k_x/2\right)+\sigma _0 \left(J_1+J_3 e^{i k_y}\right)\right)
\end{align*}

The Hamiltonian representations are given in the block form because of the choice of the representation of $S_z \sigma_3$.
It is helpful to consider how HP boson operators transform under SO(2) spin rotation.
A HP creation operator at the sublattice with $\left<S_z\right> >0$, $b_{\uparrow}^{\dagger}$ transforms as $S^{-}$ in the global spin coordinate.
On the other hand, a HP creation operator at the sublattice with $\left<S_z\right> <0$, $b_{\downarrow}^{\dagger}$  transforms as $S^{+}$ in the global spin coordinate.
Note that the arrow in subscript of $b_{\uparrow}^{\dagger}$ is to represent $\left<S_z\right>$ of the sublattice, not the spin of HP boson.
Therefore, under the rotation $b_{\uparrow}^{\dagger} b_{\downarrow}$ is not invariant while $b_{\uparrow}^{\dagger} b^{\dagger}_{\downarrow}$ is invariant.
As the result, there is no hybridization between $\left\{b_{\uparrow,-\vec{k} },\cdots,b^{\dagger}_{\downarrow \vec{k}},\cdots\right\}$ and $\left\{b^{\dagger}_{\uparrow,\vec{k}},\cdots,b_{\downarrow,-\vec{k}},\cdots\right\}$.

\end{document}